\documentclass[twocolumn]{revtex4-1}    
\usepackage[utf8]{inputenc} 
\usepackage{amsmath}
 
\usepackage{graphicx}
\begin{document}

\title{CG-content log-ratio distributions of \textit{Caenorhabditis elegans} and \textit{Drosophila
melanogaster} mirtrons}
 
\author{Denise Fagundes-Lima \footnote{defalima@gmail.com}}
\affiliation{Departamento de Ciências Biológicas, Universidade Federal de Ouro Preto,
        35400-000 Ouro Preto, MG, Brazil\\
      and Departamento de Física, Universidade Federal de Minas Gerais,31270-901 Belo Horizonte, MG, Brazil}
\author{Gerald Weber\footnote{gweberbh@gmail.com}}
\affiliation{Departamento de Física, Universidade Federal de Minas Gerais,31270-901 Belo Horizonte, MG, Brazil}
      
\begin{abstract}
Mirtrons are a special type of pre-miRNA which originate from intronic regions
and are spliced directly from the
transcript instead of being processed by Drosha.
The splicing mechanism is better understood for the processing of mRNA for which
was established that
there is a characteristic CG content around splice sites.
Here we analyse the CG-content ratio of pre-miRNAs and mirtrons and compare them
with their genomic neighbourhood
in an attempt to establish key properties which are easy to evaluate and to
understand their biogenesis.
We propose a simple log-ratio of the CG-content comparing the precursor sequence and
is flanking region.
We discovered that \textit{Caenorhabditis elegans} and \textit{Drosophila
melanogaster} mirtrons, so far without exception, have smaller CG-content than
their genomic neighbourhood. 
This is markedly different from usual pre-miRNAs which mostly have larger CG-content when
compared to their genomic neighbourhood.
We also analysed some mammalian and primate mirtrons which, in contrast
the invertebrate mirtrons, have higher CG-content ratio.
\end{abstract}

\maketitle

\section*{Introduction}

During the last decade, a wealth of small RNAs were discovered and with them new
classes of 
biological regulators emerged. 
Among those, microRNAs (or miRNAs) due to their crucial role in genomic
regulation are perhaps 
the most intensively studied.
miRNAs are involved in the regulation of numerous cellular processes including 
differentiation, development, apoptosis, proliferation, 
the stress response and they change the expression of genes in several human 
diseases such as diabetes, cancer and neuromuscular dystrophy~\cite{zhang07,iorio12,hussain12}. 

miRNAs are non-coding RNAs first identified in 1993 in the nematode
\textit{Caenorhabditis elegans}~\cite{lee93}. 
Canonical miRNAs are derived from primary miRNA transcripts (pri-miRNA), usually
long nucleotide sequences that form specific hairpin-shaped stem–loop secondary
structures. 
Pri-miRNA may originate one or more hairpins typically with 55--70 nucleotide
(nt) in length. 
In animals, pri-miRNAs are cleaved by the nuclear Drosha RNase III enzyme to release precursor miRNA (pre-miRNA) 
hairpins.
These are then transported to the cytoplasm by Exportin-5 (Exp5)
and cleaved by the Dicer RNase III enzyme to generate a very short miRNA/miRNA*
duplex~\cite{kim09}. 
One of the strands, called mature miRNA (22--25nt), is incorporated into a RISC complex (RNA induced
silencing complex) and guides the complex to the target mRNA 
to regulate gene expression while the other strand seems to take on other biological functions~\cite{kim09,okamura08,bhayani11}. 
In animals, most of the miRNA functions are related to down-regulation of genes. 

Ruby \emph{et al.}~\cite{ruby07} showed the existence of intronic pre-miRNAs in
\textit{Drosophila melanogaster} 
and \textit{C. elegans} that bypass Drosha processing providing an alternative
pathway for miRNA biogenesis~\cite{okamura07}. 
These pre-miRNAs were called 
`mirtrons' and the main difference between them and canonical miRNAs is that
intronic sequences form lariats and the 
mirtrons are originated by 
splicing~\cite{okamura07,ruby07,berezikov07}. 
Flynt \emph{et al.}~\cite{flynt10} reclassified a subset of mirtrons in
\textit{D. melanogaster} as ``tailed mirtrons", 
which have substantial $3'$ overhangs and are targets of exosome-mediated
$3'-5'$ trimming, which allows 
functional pre-miRNA to be generated. 
The existence 
of mirtrons in mammalians (human, macaque, chimpanzee, rat and/or mouse) was reported by Berezikov \emph{et
al.}~\cite{berezikov07} where they identified, using computational and
experimental strategies, 
3 well conserved mirtrons expressed in diverse mammals, 16 primate specific
mirtrons, and 46 candidate mirtrons in primates.

For mRNA, which is processed by splicing, Zhang \emph{et al.}~\cite{zhang11} 
determined that there is a characteristic CG content around splice sites.
Also, it was shown that alternative splicing is promoted by the secondary RNA
structure~\cite{shepard08} 
which is strongly determined by CG content~\cite{weber06}.
MicroRNAs are co-expressed with mRNAs~\cite{morlando08,shomron09} and, in
particular, mirtrons are seemingly not processed by the Drosha microprocessor
but by splicing only.
With splicing being dependent on thermodynamic stability could there be some
characteristic CG content which would 
set aside mirtrons from ordinary pre-miRNAs?
Of special interest would be properties which would help to understand the splicing
mechanism proposed for mirtrons~\cite{okamura07}.

Here we set out to characterise precursor sequences of miRNAs and mirtrons in
terms of CG-content and also Gibbs free energies
for \textit{D. melanogaster} and \textit{C. elegans}.
We found that the CG-content shows marked differences for both types of small
RNA.
Also, we performed the same analysis for mammalian mirtrons reported by Berezikov
\emph{et al.}~\cite{berezikov07}, again our results
show important differences albeit opposite of those for the two invertebrates.

\section*{Methods}

To characterise the small RNAs we compare the CG-content of the precursor
sequences which originate the pre-miRNAs and
mirtrons to the CG-content of their neighbouring regions.
The rationale for this approach is that if the neighbouring DNA sequence has an
important difference in thermodynamic 
stability, when compared to the precursor sequence there should be tell tale signs of it in
the CG-content fractions.
We define the CG-content fraction as
\begin{equation}
f=\frac{\text{number of C and G nucleotides}}{\text{total number of
nucleotides}}
\end{equation}
Two types of CG-content fractions are used, one $f_P$ is related to either the precursor miRNA
or the precursor mirtron. 
The other $f_N$ accounts for the total CG-contents of the 150 base pairs
downstream and upstream of the precursor sequence
which forms the neighbourhood of the precursor.
The flanking sequence length was chosen to be of the same order of magnitude of 
the length of canonical pre-miRNAs.
We perfomed the same analysis with longer flanking sequences (up to 250~nt, not presented), but found no difference from the results
reported in this work.
Both CG-contents are combined to form a log-ratio between the precursor and its
neighbourhood
\begin{equation}
R=\log_2 \left(\frac{f_P}{f_N}\right)
\end{equation}
A positive ratio means that the CG-content $f_P$ of
the precursor sequence is larger than that of its neighbours.
Since CG-content is related to thermodynamic stability we may infer that $R>0$
generally means that the flanking DNA region is less 
stable that the precursor region.
To ease the notation we use 
\begin{displaymath}
\begin{array}{cl}
R^+\rightarrow R>0 & \text{\emph{precursor} region has larger CG-content}, f_P>
f_N\\
R^-\rightarrow R<0 & \text{\emph{flanking} region has larger CG-content}, f_N>
f_P
\end{array}
\end{displaymath}

To evaluate the statistical significance of our findings we use 
the Kolmogorov-Smirnov test \cite{press92}.
Even though this statistical test is well established, given the question which is posed in
this work it is perhaps more intuitive and simpler to quantify the significance by
using simple combinatorial probabilities.
Therefore, we also calculate the probability $p^-$ of drawing $k$ pre-miRNAs, all with $R^-$, purely by chance 
\begin{equation}
p^-=\left(\frac{n^-}{n^-+n^+}\right)^{k}
\label{eq-prob}
\end{equation}
where $n^-$ and $n^+$ are number of known pre-miRNAs with $R^-$ and $R^+$,
respectively.

The database used to obtain the precursor miRNA and mirtrons of \textit{D.
melanogaster} and \textit{C. elegans} was from mirBASE version 16
\cite{griffiths-jones06,kozomara11}, 
which is one of the main on-line repositories for microRNA sequences. 
For extracting the flanking sequences we used the complete genome file of
\textit{D. melanogaster} 
version r5.34 \cite{drysdale05} and version WS223 for \textit{C. elegans}
\cite{stein01}.
We retrieved the precursor miRNA/mirtron sequences by searching for an exact match within
the complete genome files. 
For each sequence four types of matches were performed: the original sequence, 
the reversed sequence, the complementary sequence and the reversed-complementary sequence.

The mirtrons reported by  Berezikov \emph{et al.}~\cite{berezikov07} were
collected 
from the supplemental data, the neighbouring sequences for each these mirtrons were
obtained from Ensemble API and databases~\cite{hubbard02}.

To complete our analysis we also calculated the average Gibbs free energies of mirtrons and ordinary pre-miRNA.
In this work we use the RNAfold program from the Vienna package
\cite{hofacker03} with default parameters to obtain the Gibbs free energies, $\Delta G$.

\section*{Results and Discussion}

\begin{figure}[htpb]
\begin{center}
\includegraphics[width=0.4\textwidth]{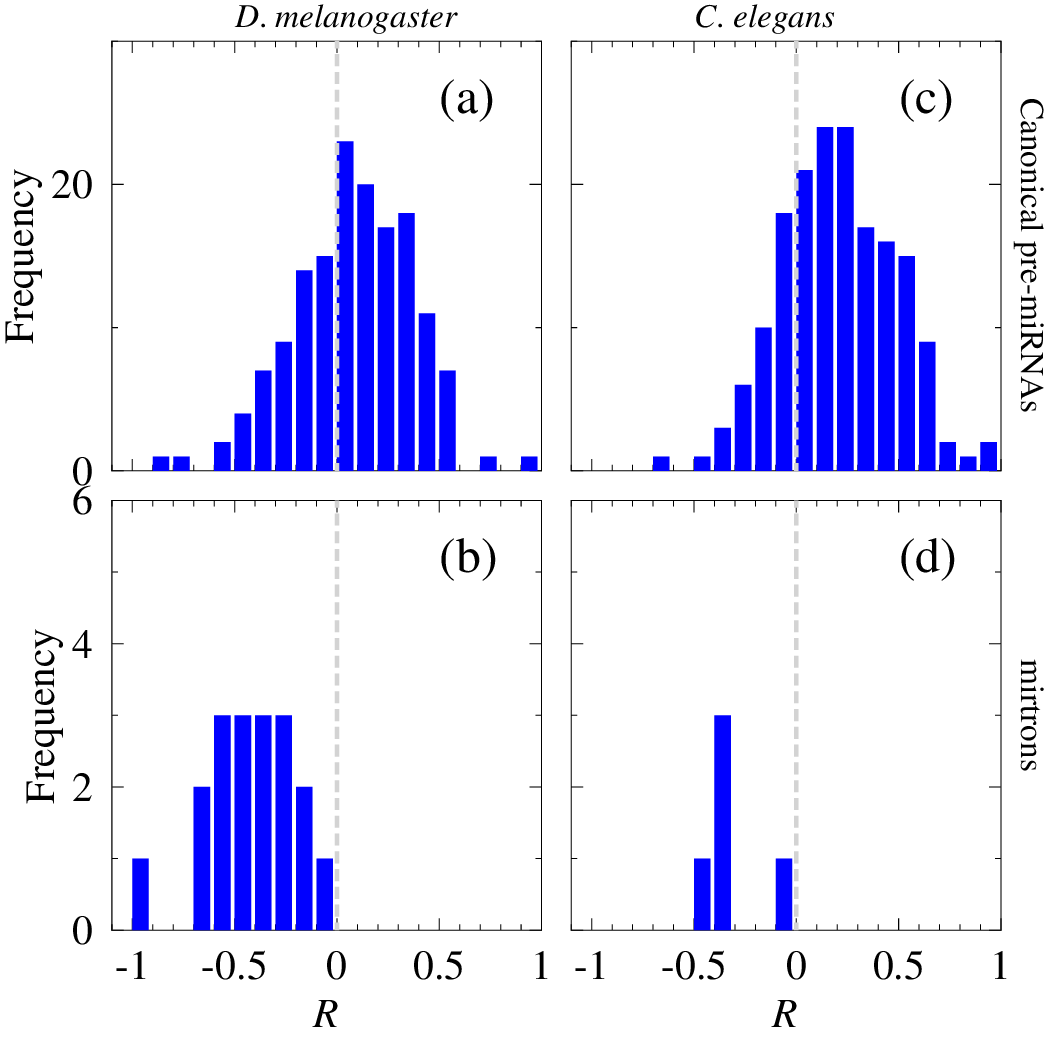}
\end{center}
\caption{CG-content ratio~$R$ distribution for a)~151 canonical miRNAs and b)~18
mirtrons 
of \textit{D. melanogaster}
and  c)~170 canonical miRNAs and d)~5 mirtrons of \textit{C. elegans}.}
\label{fig:R}
\end{figure}

\begin{table*}[htbp]
\caption{\textbf{CG-content ration~$R$ and free energy~$\Delta G$ characteristics and of canonical pre-miRNAs and 
mirtrons of invertebrates. 
Also shown are number of sequences $n^\pm$ with $R^\pm$, average CG-content ratio~$\langle R\rangle$, 
average free energy $\langle\Delta G\rangle$  and
average precursor length $\langle N\rangle$.}}
\label{Tab:cel-dmel}
\footnotesize
\begin{tabular}{lllrrcrcr}\hline\hline
Organism & RNA type & total & $n^+$ & $n^-$ & $n^+:n^-$ & $\langle R\rangle$ & $\langle\Delta G\rangle$ (kcal/mol) & $\langle N\rangle$ (nt)\\
\textit{C. elegans} & mirtrons&5 &0 & 5 &  & $0.81 \pm0.08$ & $-20.28 \pm 4.21$ & $62.22\pm6.72$\\
\textit{D. melanogaster}&mirtrons&18 & 0 & 18 &  & $ 0.76 \pm0.10$ & $-22.03 \pm 6.55$ & $69.05 \pm16.68$\\
\textit{D. melanogaster} &tailed mirtrons &7 &2 & 5 & $1:1.8$ & $0.84 \pm0.24$ & $-20.36 \pm 10.19$ & $93.00\pm42.24$\\
\textit{C. elegans}  &canonical miRNAs\footnotemark[1]&170 &131 & 39 & $3.3:1$& $1.18 \pm0.23$ & $-35.10 \pm 9.09$ & $91.78\pm14.26$\\
\textit{D. melanogaster}  &canonical miRNAs\footnotemark[1] &151 & 97 & 54 & $1.8:1$ & $1.08 \pm0.21$ & $-33.89 \pm 8.25$ & $94.00\pm18.50$\\
\hline\hline
\end{tabular}\\
{\footnotemark[1]mirtrons excluded.}
\caption{\textbf{CG-content ratio~$R$ and free energy~$\Delta G$ characteristics and of canonical 
miRNAs considering only those with $R^-$.}
\label{Tab:negative}}
\begin{tabular}{llrrrc}\hline\hline
Organism & RNA type & $n^-$ & $\langle R^-\rangle$ & $\langle\Delta G\rangle$ (kcal/mol)& $\langle N\rangle$ (nt)\\
\textit{C. elegans} & canonical miRNAs &39  &  $0.90 \pm0.08$ & $-30.17 \pm8.72$ & $90.95\pm17.50$\\
\textit{D. melanogaster}&canonical miRNAs & 54 & $0.87 \pm0.10$ & $-30.0 \pm 5.40$ & $90.40 \pm11.78$\\
\hline\hline
\end{tabular} { }
\end{table*}

In Fig.~\ref{fig:R} we show the distribution of CG-content log-ratio~$R$ for
canonical pre-miRNAs and mirtrons, defined in Methods, 
of \textit{D. melanogaster} and \textit{C. elegans}.
The content log-ratio~$R$ for canonical pre-miRNAs, Fig.~\ref{fig:R}a, is roughly
gaussian with a peak around $R=0$.
This means that for this type pre-miRNA there appears to be no strong preferential
ratio for CG-content within the precursor
sequence and its neighbours, although a bias towards $R^+$ is clearly noticeable.
In stark contrast, all 18~mirtrons of \textit{D. melanogaster} have $R<0$
($R^-$) as shown in Fig.~\ref{fig:R}b.
Even though the number of reported mirtrons is still small, 
the probability of picking 18 small RNAs with $R^-$ by chance alone, considering the
distribution for canonical pre-miRNA,
is $p^-=9.1\times10^{-9}$, see Eq.~(\ref{eq-prob}).
The Kolmogorov-Smirnov distibution test yields $p^-=7.8\times10^{-9}$ which essentially confirms
the simple combinatorial probability.
Therefore, the occurrence of 18 $R^-$ pre-miRNA entirely by chance is very unlikely.

Some authors describe mirtrons as tightly packed between exons~\cite{hussain12}, 
but in our analysis we have found that this is not the case. 
Most mirtrons are surrounded by intronic sequences not exons.
This seems consistent if one considers that intronic regions of \textit{D. melanogaster} are about 750 to 1000~nt 
in length on average~\cite{presgraves06} and that mirtrons are typically 60~nt in length.
Therefore, $R^-$ means that the immediate flanking region which is also intronic is
more stable than the precursor region.
One possible explanation for the predominance of $R^-$ would be if intronic
regions were of highest CG-content.
However, the intronic regions of \textit{D. melanogaster} have one of the smallest
CG-content in this genome: 0.4 as compared to 0.52 for coding regions.
The fact that the surrounding region of mirtrons has a higher CG content, which is unusual for intronic regions,
suggest a role in the processing of these special types of miRNAs.
Therefore, we may speculate that $R^-$ may play a role in the mirtron splicing mechanism in a similar fashion to
what happens for messenger RNA~\cite{zhang11}.

For canonical \textit{C. elegans} pre-miRNAs we observe a similar gaussian
shaped distribution of the ratio~$R$
(Fig.~\ref{fig:R}c) but with strong bias towards  $R^+$.
Tab.~\ref{Tab:cel-dmel} shows that the $n^+:n^-$ ratio 
is of three $R^+$ pre-miRNAs for each $R^-$ pre-miRNA.
To date there are only five mirtrons reported and they
all show $R^-$ (Fig.~\ref{fig:R}d), similar to the mirtrons of \textit{D. melanogaster}.
Even though this number is very small, it is still intriguing given the strong bias toward $R^+$
in canonical pre-miRNAs. 
Indeed, the probability of picking 5 pre-miRNAs all with $R^-$ is small,
the combinatorial probability being $p^-=6.3\times10^{-4}$. 
Again, the Kolmogorov-Smirnov test provides $5.5\times10^{-4}$ in agreement with the combinatorial probability. 

\begin{figure}[htpb]
\begin{center}
\includegraphics[width=0.4\textwidth]{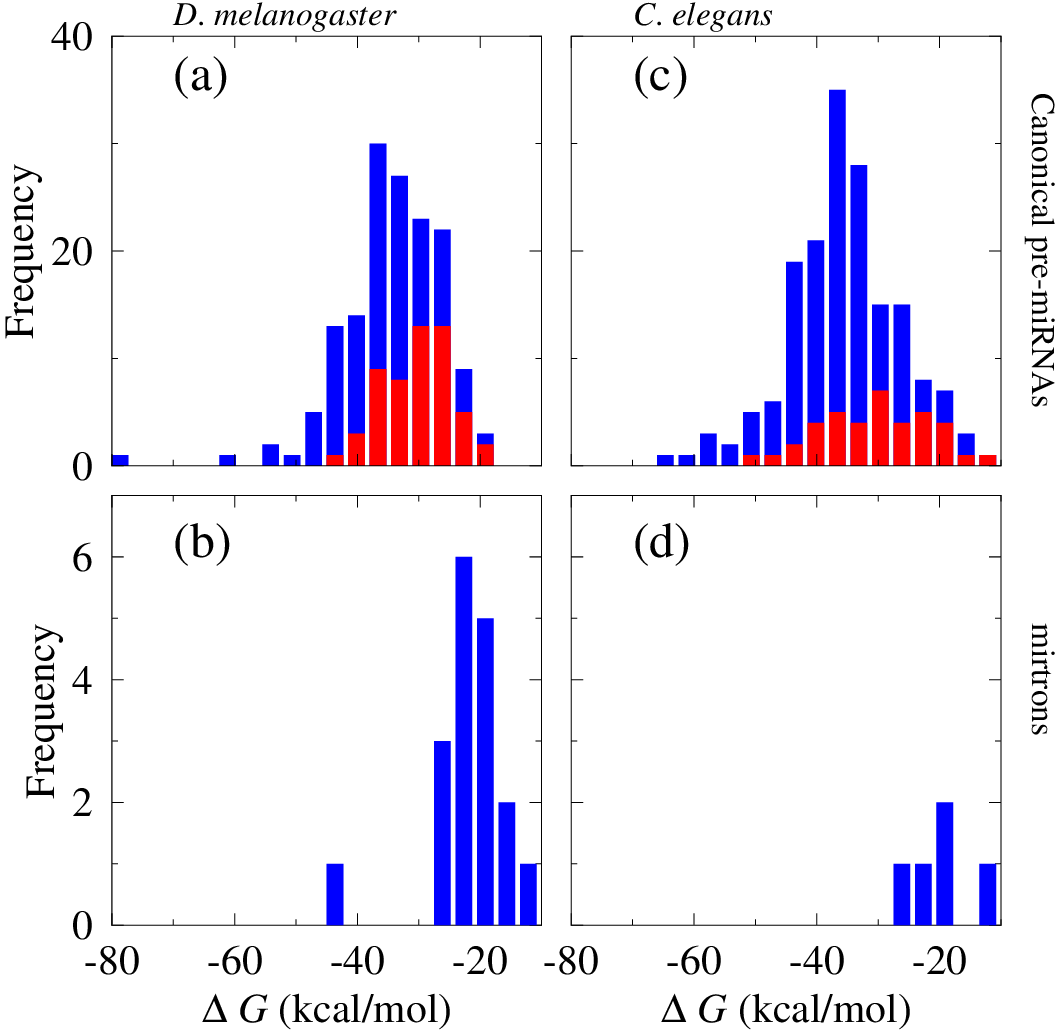}
\end{center}
\caption{Average free energy ~$\Delta G$ distribution for a)~151
canonical pre-miRNAs and b)~18 mirtrons of \textit{D. melanogaster}.
and  c)~170 canonical miRNAs and d)~5 mirtrons of \textit{C. elegans}. Red bars
are for $R^-$ miRNAs.}
\label{fig:DG}
\end{figure}

To complete our comparative analysis of mirtrons and canonical pre-miRNAs, we also 
calculated the Gibbs free energies.
Clearly, given the $R^-$ nature of the mirtons, one would expect these to be generally less stable
than the average pre-miRNAs.
In Fig.~\ref{fig:DG} we show the distribution of free energy 
$\Delta G$ for both invertebrates, and detailed quantities are also
given in Tab.~\ref{Tab:cel-dmel}.
Except for one notable exception, all mitrons show $\Delta G$ larger than $-30$~kcal/mol, confirming their instability.
In contrast, canonical pre-miRNAs are distributed over a much larger range of energies.
Certainly, the fact that mirtrons are much shorter than canonical pre-miRNAs, 60~nt compared to 90~nt on average,
largely accounts for this.
But is a free energy  larger than $-30$~kcal/mol sufficient to result in $R^-$?
To answer this, we isolated all canonical pre-miRNAs with $R^-$ and recalculated
the their $\Delta G$ distribution, which are shown 
as red bars in Figs.~\ref{fig:DG}a and~\ref{fig:DG}c and summarised in Tab.~\ref{Tab:negative}.
Essentially, we find a considerable number of $R^+$ pre-miRNAs with $\Delta G > -30$~kcal/mol.
In other words, a pre-miRNAs with $\Delta G > -30$~kcal/mol does not imply in $R^-$.
Therefore, the free energy distribution alone does not explain why all mirtrons of 
\textit{D. melanogaster} and \textit{C. elegans}
are $R^-$.

\begin{figure}[htbp]
\begin{center}
\includegraphics[width=0.4\textwidth]{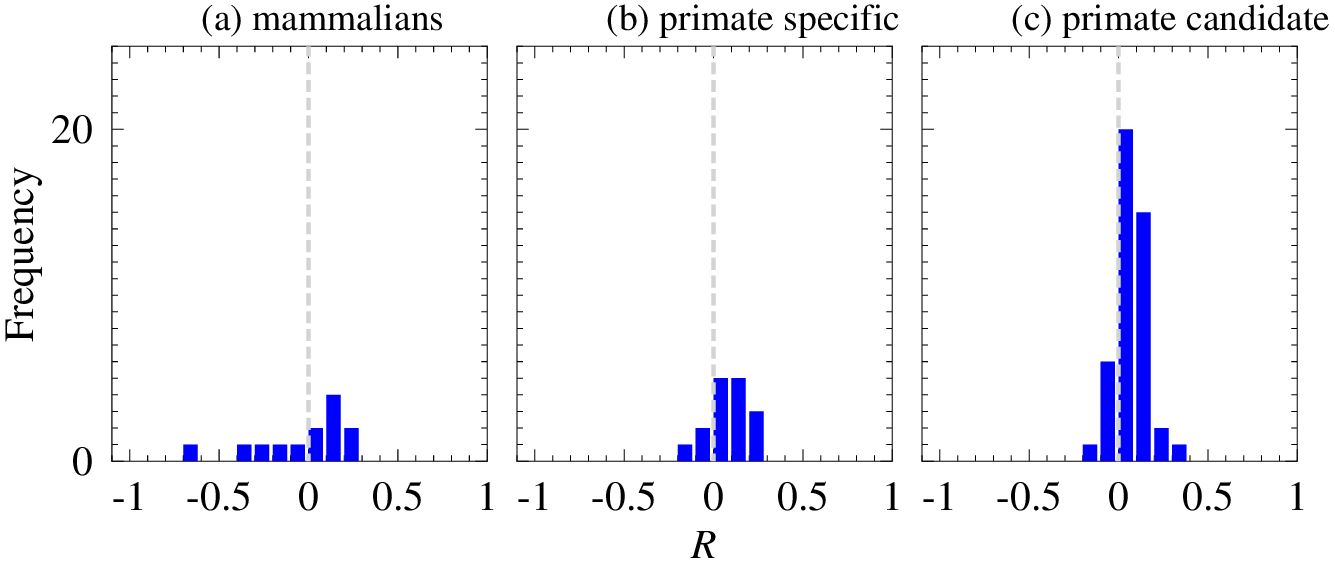}
\end{center}
\caption{CG-content ratio~$R$ distribution for a)~13 putative mammalian
mirtrons,
b)~16 specific primate mirtrons and c) 46 candidate primate mirtrons.}
\label{fig:sup1}
\end{figure}

The next question is whether other types of reported mirtrons, such as primate
and mammalian mirtrons
show the same $R$ distribution as \textit{D. melanogaster} and \textit{C.
elegans}?
As shown in Tab.~\ref{Tab:vert},
in terms of CG-content ration~$R$ and average free energy $\Delta
G$ these mirtrons appear not to be biased to any particular value. 
Berezikov \emph{et al.}~\cite{berezikov07} found that the GC content of mammalian mirtrons was much higher than that 
of invertebrate miRNAs but,
in comparison with their neighbours regions, 
we found that they tend generally to $R^+$ (\text{\emph{precursor} region has larger CG-content}), 
see Fig.~\ref{fig:sup1}.
We have not attempted to generate the distribution of mammalian pre-miRNAs due to the number of large genomes
which would have to be processed.

\begin{table*}[bhtp]
\caption{\textbf{CG-content ratio~$R$ and free energy~$\Delta G$ characteristics of specific vertebrate mirtrons.
\label{Tab:vert}}}
\small
\begin{tabular}{lllrrcrrc}\hline\hline
Organism & RNA type & total &  $n^+$ & $n^-$ & $n^+:n^-$ & $\langle R\rangle$ & $\Delta G$ (kcal/mol)& $\langle N\rangle$ (nt)\\
mammalians& putative mirtrons &13 & 8 & 5 & $1.5:1$ & $ 0.98 \pm0.16$ & $-30.02 \pm 13.20$ & $87.92 \pm9.52$\\
primates  & specific mirtrons  &16 & 13 & 3 & $4.3:1$ & $1.12\pm0.11$ & $-30.32 \pm11.69$ & $83.62 \pm24.36$\\
primates  &  candidate mirtrons &45\footnotemark[2] &40& 5 & $8:1$ & $1.11 \pm0.09$ & $-41.18 \pm12.39$ & $90.44\pm21.94$\\
\hline\hline
\end{tabular}\\
{\footnotemark[2]Ref.~\onlinecite{berezikov07} reports 46 candidates mirtrons
for primates, yet 
supplementary tables only show 45 sequences.}
\end{table*}

\section*{Conclusions}

We have introduced the concept of CG-content log-ratio of precursor sequences and
flanking regions and
discovered that all \textit{D.
melanogaster} and \textit{C. elegans} mirtrons are $R^-$.
This cannot be explained by the CG-content of the intronic region and neither by the fact
that mirtrons are generally shorter and less stable than pre-miRNAs.
Usual pre-miRNAs of these organisms only show a moderate bias towards $R^+$.
This finding appears to support
the notion that mirtrons are spliced in a similar fashion to mRNA instead of
being processed by Drosha.
For mammalian mirtrons we have found no such bias, but we noticed that these also display several 
important differences when compared
to the vertebrate mirtrons which were considered in this work, such as differences in length and free energy.

\section*{Acknowledgements}

We are grateful to J. M. Ortega for helpful suggestions.
Funding: CNPq, Fapemig and National Institute of Science and
Technology for Complex Systems.

 \bibliographystyle{nature}  
 \bibliography{complete-gbc,artigo}

\end{document}